\input{aipcheck}

\documentclass[
    ,final            % use final for the camera ready runs
  ]
  {aipproc}

\layoutstyle{6x9}

\begin{document}

\title{Gluon polarization measurements with inclusive jets at STAR}


\author{Pibero Djawotho for the STAR Collaboration}{
  address={Cyclotron Institute, Texas A\&M University, College Station, TX 77843-3366, USA}
}

\begin{abstract}
At RHIC kinematics, polarized jet hadroproduction is dominated by $gg$
and $qg$ scattering, making the jet double longitudinal spin asymmetry,
$A_{LL}$, sensitive to gluon polarization in the nucleon. I will present
STAR results of $A_{LL}$ from inclusive jet measurements for the RHIC
2006 run at center-of-mass energy 200 GeV. I will
also discuss the current status of the analysis of data from the
2009 run, also at center-of-mass energy 200 GeV.
The results are compared with theoretical calculations of $A_{LL}$ based
on various models of the gluon density in the nucleon. The STAR data
place significant constraints on allowed theoretical models.
\end{abstract}

\maketitle

%%%%%%%%%%%%%%%%%%%%%%%%%%%%%%%%%%%%%%%%%%%%
%% MAINMATTER
%%%%%%%%%%%%%%%%%%%%%%%%%%%%%%%%%%%%%%%%%%%%

\section{Introduction}

The study of the internal spin structure of the proton is an integral
part of the Relativistic Heavy Ion Collider (RHIC). The polarized
proton collider is especially well suited to measure the polarized
gluon contribution to the proton spin. The Solenoidal Tracker at
RHIC (STAR) experiment, with its large acceptance, retains an advantage
in accessing $\Delta g(x)$ via jet production.

The longitudinal spin sum rule dictates how the proton spin is constructed
from the spin and orbital angular momenta of its partonic constituents:
\begin{equation}
\frac{1}{2}=\frac{1}{2}\Delta\Sigma+\Delta G+L_z,
\end{equation}
where the quark polarization, $\Delta\Sigma \approx 0.3,$ has been measured
in deep-inelastic scattering experiments. However, $\Delta G,$ the gluon
polarization, and $L_z,$ the parton orbital angular momentum, are still
poorly constrained \cite{Ashman}. RHIC stands to bring significant advances
in mapping $\Delta g(x)$.

RHIC collected data at 200 GeV center-of-mass energy in polarized
proton-proton collisions with integrated luminosity of 5.4 pb$^{-1}$
in 2006 and 25 pb$^{-1}$ in 2009. The beam polarization was measured
with Coulomb-nuclear interference (CNI) proton-carbon polarimeters \cite{Jinnouchi}
calibrated with a polarized atomic hydrogen gas-jet target \cite{Okada}.
The average beam polarization was 60\% in 2006 and 58\% in 2009.

The STAR detector subsystems \cite{Harrison} relevant to jet analysis are
the Time Projection Chamber (TPC) immersed in a 0.5 T longitudinal
magnetic field and used to reconstruct charged particle tracks with pseudorapidity
$|\eta|<1.3$, the Barrel Electromagnetic Calorimeter (BEMC)
with towers at $|\eta|<1$ and the Endcap Electromagnetic Calorimeter
(EEMC) with towers at $1<\eta<2$. The Beam-Beam Counters (BBC) with
$3.3<|\eta|<5.0$ and Zero-Degree Calorimeters (ZDC) located
$\sim$18 m downstream of the interaction point were used for
monitoring relative luminosities. A timing window imposed on the
BBCs was used as part of the minimum bias trigger requirement in 2006.
All of these detectors cover full azimuth ($\Delta\phi=2\pi$).

\section{Analysis and results}

\begin{figure}
\includegraphics[width=0.5\textwidth]{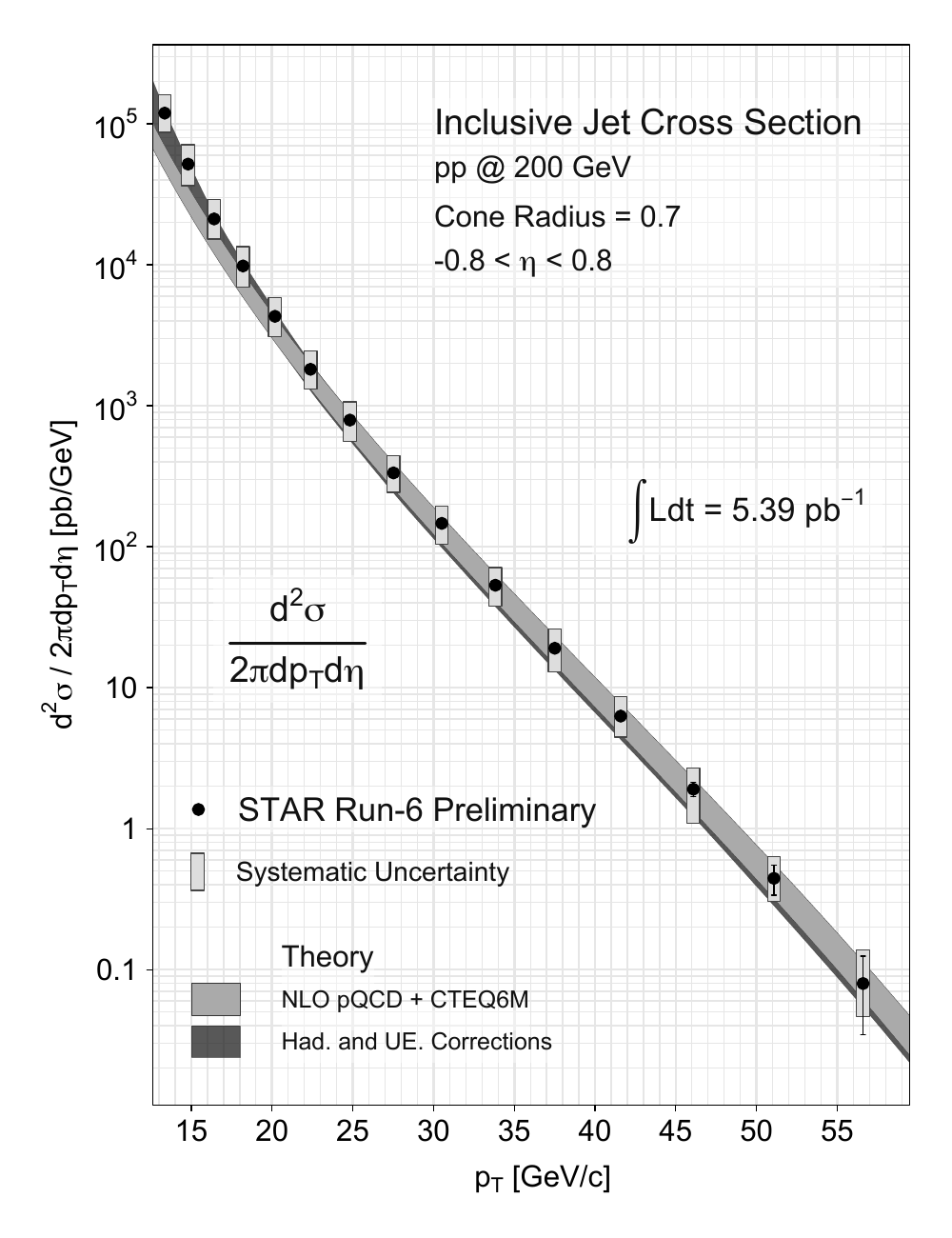}
\includegraphics[width=0.5\textwidth]{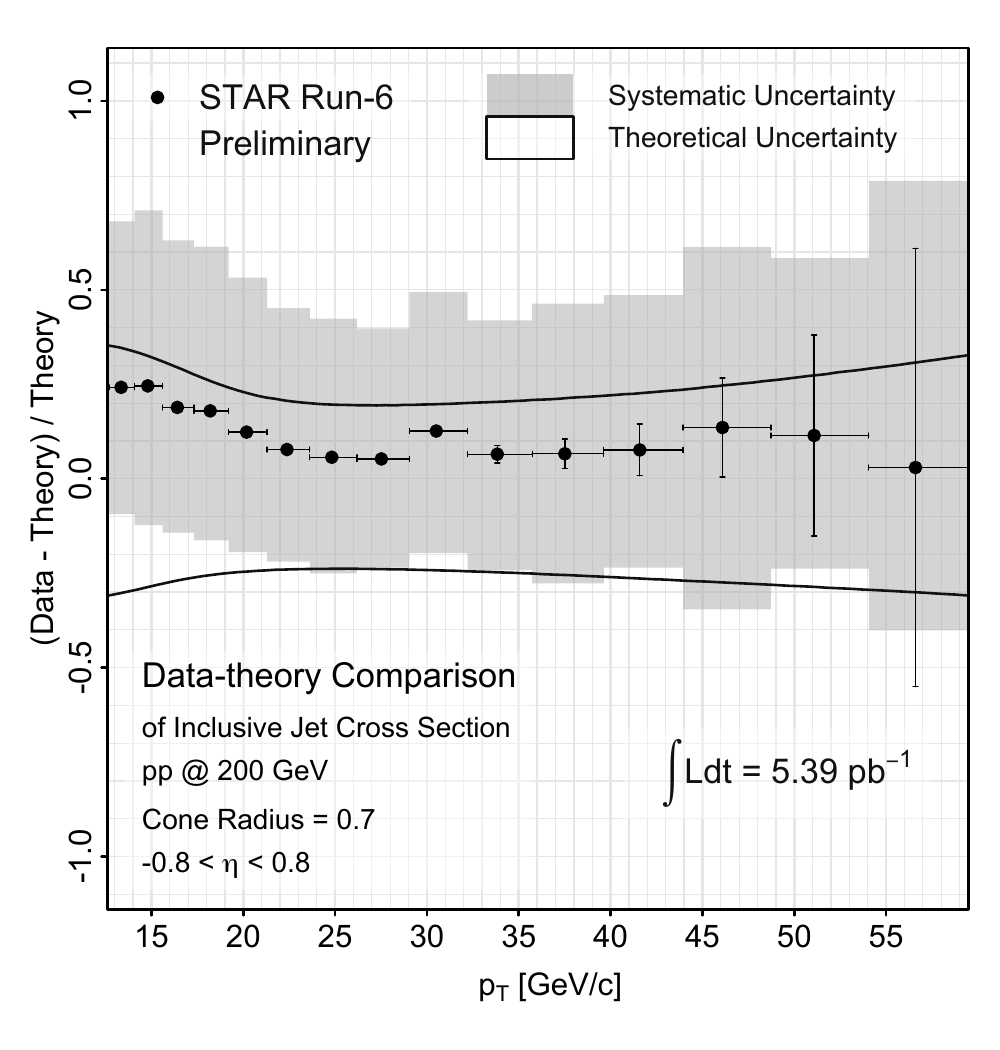}
\caption{2006 inclusive jet cross section (left panel) and (data-theory)/theory
comparison (right panel) versus jet $p_T$}
\label{fig:jetCrossSection}
\end{figure}

Jets were reconstructed using a midpoint-cone algorithm \cite{Blazey}
with seed transverse energy 0.5 GeV, split-merge fraction 0.5 and
cone radius 0.7. Tracks and towers were required to have a minimum
transverse momentum of 0.2 GeV/$c$. The inclusive jet differential
cross section was calculated from the data with \cite{Abelev}:
\begin{equation}
\frac{1}{2\pi}\frac{d^2\sigma}{d\eta dp_T}=\frac{1}{2\pi}
\frac{N_{jets}}{\Delta\eta\Delta p_T}\frac{1}{\int L dt}\frac{1}{c(p_T)},
\end{equation}
where $N_{jets}$ is the number of jets in the bin $(\Delta\eta, \Delta p_T)$
and $c(p_T)$ is a correction factor determined from simulation
with the event generator PYTHIA 6.4 \cite{PYTHIA} with CDF Tune A settings
and GEANT \cite{GEANT} detector response. Figure~\ref{fig:jetCrossSection}
shows the 2006 measured inclusive jet cross section with theory comparisons
\cite{Jager,Pumplin}. With the inclusion of hadronization and underlying event
corrections, the STAR data are well described by theory.

Several important improvements over the 2006 run were realized,
both before and after the taking of the 2009 data. Overlapping
jet patches were added to the trigger and lower $E_T$ thresholds
were adopted for both the BEMC and EEMC. These upgrades helped
increase trigger efficiency and reduce trigger bias. They resulted
in a 37\% increase in jet acceptance over the 2006 run. Upgrades
in the data acquisition system, DAQ1000, allowed STAR to record
events at several hundred Hz during the 2009 run, with only 5\%
dead time for the jet data, compared with 40 Hz with 40\% dead
time during the 2006 run. The enhanced DAQ capability also
allowed STAR to remove the BBC coincidence requirement, which
helped trigger more efficiently at high jet $p_T$ and perform
the first direct measurement of non-collision background at STAR.
Improvements in jet reconstruction were also implemented. The
electromagnetic calorimeters are $\sim$1 hadronic interaction length
thick. Many charged hadrons deposit a MIP (minimum ionizing particle),
while others shower and deposit a sizeable fraction of their energy
when passing through. The strategy adopted in analyses through 2006
was to subtract a MIP from an EMC tower with a charged track passing
through. In the 2009 run, the total momentum of the charged track
is subtracted from the struck EMC tower. This significantly reduces
the response to fluctuations from charged hadron showering and
reduces the average difference between jet energies at the detector
and particle level. The net benefit comes in the form of an improved
overall jet energy resolution of 18\%, compared to 23\% in the 2006
anaysis.

\begin{figure}
\includegraphics[height=.294\textheight]{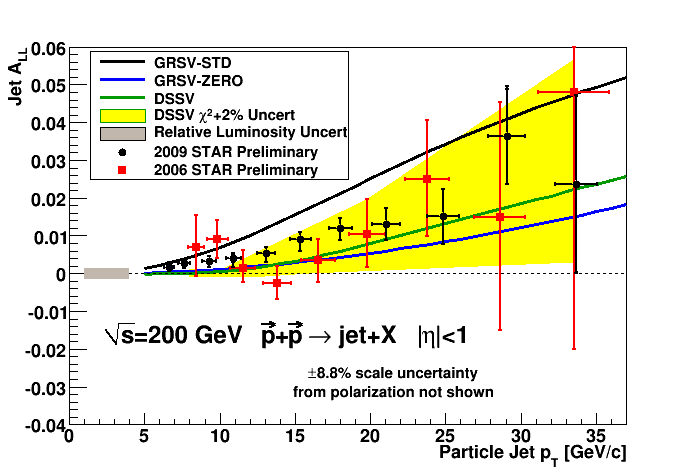}
\caption{2006 (red squares) and 2009 (black circles) inclusive jet $A_{LL}$
versus jet $p_T$}
\label{fig:jetALL}
\end{figure}

\begin{figure}
\includegraphics[width=0.5\textwidth]{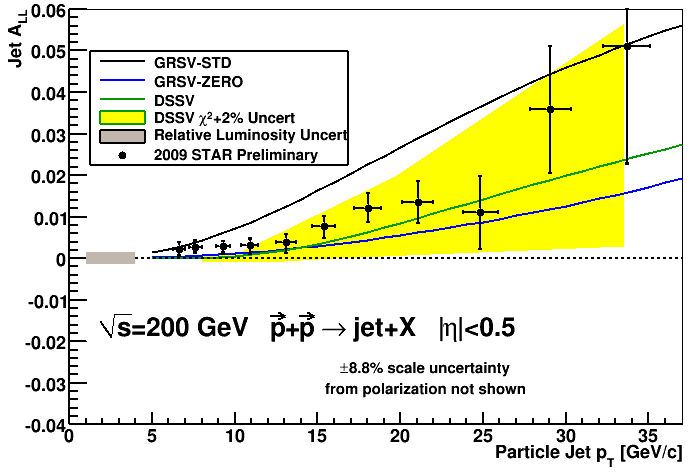}
\includegraphics[width=0.5\textwidth]{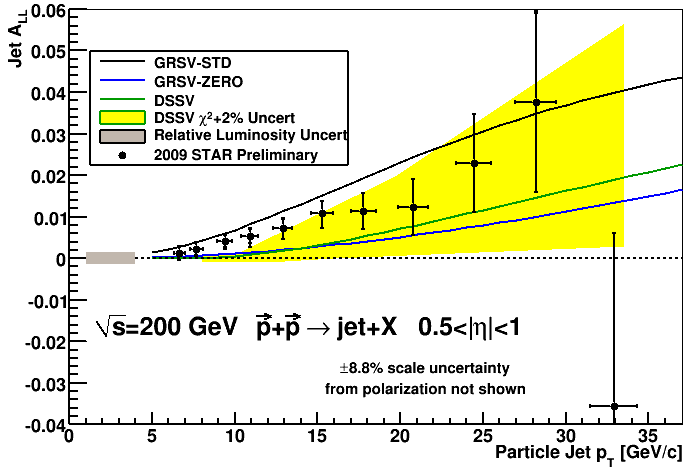}
\caption{2009 inclusive jet $A_{LL}$  versus jet $p_T$ for the pseudorapidity ranges
$|\eta|<0.5$ (left panel) and $0.5<|\eta|<1$ (right panel)}
\label{fig:jetALL2ranges}
\end{figure}

Figure~\ref{fig:jetALL} shows the measured inclusive jet
$A_{LL}=(\sigma^{++}-\sigma^{+-})/(\sigma^{++}+\sigma^{+-})$
versus jet $p_T$ for the 2006 ($-0.7<\eta<0.9$) and 2009 ($|\eta|<1$)
data alongside theory predictions of GRSV \cite{GRSV} and DSSV \cite{DSSV}. 
The STAR data fall between the predictions of the two models.
The dominant systematic uncertainties originate from differences
between the reconstructed and true jet $p_T$ and the trigger sampling
the underlying partonic processes ($qq$, $qg$ and $gg$) differently.
The 2009 data are more precise than the 2006 data by a factor of four
in low-$p_T$ bins and a factor of three in high-$p_T$ bins.
Figure~\ref{fig:jetALL2ranges} presents the 2009 result in two rapidity ranges,
permitting comparisons with models for collisions with different average
partonic scattering angles, $x$ ranges, and subprocess mixtures.

\section{Conclusion}

The STAR experiment measured the inclusive jet differential cross
section and longitudinal double-helicity asymmetry $A_{LL}$ in
polarized proton-proton collisions at $\sqrt{s}=200$ GeV. The agreement
between NLO pQCD calculations and the measured cross section runs over
six orders of magnitude for $p_T\sim$13--57 GeV/$c$. The STAR 2005 and 2006
measured inclusive jet $A_{LL}$ were included in the first global analysis
\cite{DSSV} to use polarized jets and played a significant role in
constraining $\Delta g(x)$ at RHIC kinematics. The markedly increased
precision of the 2009 result is expected to vastly reduce
the present large uncertainty of the gluon polarization of the proton
once included in a global analysis of polarized parton densities.
In addition, the complementary STAR measurement of $A_{LL}$ for di-jets
for 2009, also presented for the first time at this meeting \cite{Walker},
will reduce the uncertainty in $\Delta g(x)$ associated with extrapolating
beyond the $x$ range explored by the inclusive jet measurement.

%%%%%%%%%%%%%%%%%%%%%%%%%%%%%%%%%%%%%%%%%%%%%%%%
%% BACKMATTER
%%%%%%%%%%%%%%%%%%%%%%%%%%%%%%%%%%%%%%%%%%%%%%%%



\bibliographystyle{aipproc}   % if natbib is available

%\bibliography{sample}

\begin{thebibliography}{99}

\bibitem{Ashman}
J.~Ashman {\it et al.}, Nucl. Phys. {\bf B 328}, 1 (1989);
B.W.~Filipone and X.D.~Ji, Adv. Nucl. Phys. {\bf 26}, 1 (2001).

\bibitem{Jinnouchi}
O.~Jinnouchi {\it et al.}, arXiv:nucl-ex/0412053.

\bibitem{Okada}
H.~Okada {\it et al.}, arXiv:hep-ex/0601001.

\bibitem{Harrison}
Special Issue on RHIC and its Detectors, edited by Michael Harrison,
Thomas Ludlam, and Satoshi Ozaki [Nucl. Instrum. Methods Phys. Res.,
Sect. {\bf A 499}, 624 (2006)].

\bibitem{Blazey}
G.C.~Blazey {\it et al.}, arXiv:hep-ex/0005012.

\bibitem{Abelev}
B.I.~Abelev {\it et al.}, Phys. Rev. Lett. {\bf 97}, 252001 (2006).

\bibitem{PYTHIA}
T.~Sjostrand, L.~Lonnblad, and S.~Mrenna, arXiv:hep-ph/0108264.

\bibitem{GEANT}
GEANT 3.21, CERN Program Library.

\bibitem{Jager}
B.~J\"ager, M.~Stratmann, and W.~Vogelsang, Phys. Rev. {\bf D 70}, 034010 (2004).

\bibitem{Pumplin}
J.~Pumplin {\it et al.}, JHEP {\bf 07}, 012 (2002).

\bibitem{GRSV}
M.~Gl\"uck, E.~Reya, M.~Stratmann, and W.~Vogelsang, Phys. Rev. {\bf D 63}, 094005 (2001).

\bibitem{DSSV}
D.~de~Florian {\it et al.}, Phys. Rev. Lett. {\bf 101}, 072001 (2008).

\bibitem{Walker}
M.~Walker for the STAR Collaboration, these proceedings.

\end{thebibliography}

%  \typeout{******************************************}
%  \typeout{** Please run "bibtex \jobname" to optain}
%  \typeout{** the bibliography and then re-run LaTeX}
%  \typeout{** twice to fix the references!}
%  \typeout{******************************************}

%\end{document}

%%%%%%%%%%%%%%%%%%%%%%%%%%%%%%%%%%%%%%%%%%%
%% The following lines show an example how to produce a bibliography
%% without the help of the BibTeX program. This could be used instead
%% of the above.
%%%%%%%%%%%%%%%%%%%%%%%%%%%%%%%%%%%%%%%%%%%

\end{document}